\documentclass[%
preprint,
preprintnumbers,
 amsmath,amssymb,
 aps,
floatfix,
]{revtex4-1}

\usepackage{graphicx}
\usepackage{dcolumn}
\usepackage{bm}
\usepackage{color}
\usepackage{xspace}

\begin{document}

\preprint{CTH/2017 - 1}

\title{Self-organisation of random oscillators with L\'{e}vy stable distributions}

\author{Sara Moradi$^1$}
\email{smoradi@ulb.ac.be}
 \author{Johan Anderson$^2$}%

\affiliation{%
$^1$ Fluid and Plasma Dynamics, Universit\'{e} Libre de Bruxelles, 1050-Brussels, Belgium\\
$^2$ Department of Earth and Space Sciences, Chalmers University of Technology, SE-412 96 G\"{o}teborg, Sweden}

\begin{abstract}
{\bf Abstract}: A novel possibility of self-organized behaviour of stochastically driven oscillators is presented. It is shown that synchronization by L\'{e}vy stable processes is significantly more efficient than that by oscillators with Gaussian statistics. The impact of outlier events from the tail of the distribution function was examined by artificially introducing a few additional oscillators with very strong coupling strengths and it is found that remarkably even one such rare and extreme event may govern the long term behaviour of the coupled system. In addition to the multiplicative noise component, we have investigated the impact of an external additive L\'{e}vy distributed noise component on the synchronisation properties of the oscillators. 
\end{abstract}

\pacs{Valid PACS appear here}
\maketitle


\section{Introduction:} A simple mathematical but yet powerful tool to study the dynamics of a many-body interacting system is the Kuramoto's system of randomly coupled limit cycle oscillators. Over the years, many aspects of the model, including applications cutting across disciplines, from physical and biological to even social modelling, have been considered in the literature \cite{kuramoto, strogatz,Acebron2005,winfree, daido1, daido2, crawford, daido3, strogatz1, hong, sonnenschein}. A particular application of the model has been in turbulence theory where the model can be employed to examine various aspects of the non-linear dynamics. In a previous work we developed a predator-prey model of dual populations with stochastic oscillators to examine several important features of the dynamical interplay between the drift wave and zonal flow turbulence in magnetically confined plasmas \cite{moradi1}. The underlying reasoning was that by rewriting the function representing the fluctuation quantities as $f_k = |f_k| exp (i \theta_k (t) + i \vec{k} \cdot \vec{r})$ and following the typical quadratically nonlinear primitive equations that arise in practice, the phase evolution equation can be written as:
\begin{eqnarray}
\frac{df_k}{dt} + i\omega_{k} f_{k} + \sum_{k=k'+k''} M_{k'k''} f_{k'} f_{k''} =0 \label{1}
\end{eqnarray}
which is an equation of the Kuramoto form. By including an additive noise term on the RHS of the above equation we can pave the way for the use of efficient analytical tools commonly employed in the study of the statistical behaviour of many-body systems. Furthermore, by treating the $M_{k'k''}$ as a random coefficient we can obtain insights into properties of multiplicative statistics, albeit with radical simplifications, that are more common in practice e.g. the advective nonlinearities in the Navier-Stokes, MHD, gyrokinetic, and other equations. Thus, statistical theories can be viewed as reduced descriptions of the wealth of information in the true turbulent dynamics, and there has been a wide range of applications based on these statistical treatments e.g Langevin equations using additive forcing \cite{Langevin1908} and stochastic oscillator model employing multiplicative forcing \cite{Kraichnan79}. The latter model played an important role in the development of statistical closure techniques. 

An important shortcoming here is that the main body of work has made Gaussian statistical assumptions. Although Gaussian statistics can in certain cases of diffusion in time and space give a good representation of the apparent randomness, in many systems there are processes where the Gaussian approach is inappropriate  \cite{klafter2005,metzler2000,montroll1973,zaslovsky2,kou2004,montroll}. There is a wealth of experimental and numerical evidence that indicate that turbulent transport under some conditions, is non-diffusive \cite{ida}. There are several reasons for the possible breakdown of the standard diffusion paradigm which is based on restrictive assumptions including locality, Gaussianity, lack of long range correlations, and linearity. Different physical mechanisms can generate situations where e.g., locality and Gaussianity may be incorrect assumptions for understanding transport. For example, interactions with external fluctuations may introduce long-range correlations and/or anomalously large particle displacements \cite{moradi2,anderson1}. The source of the external fluctuations could be that not all relevant physics is taken into account, such as coherent modes or other non-linear mechanisms may be neglected. In addition, turbulence intermittency is characterized by patchy spatial structures that are bursty in time, and coupling to these modes introduces long range correlations and/or L\'{e}vy distributed noise characteristics \cite{levy}. The probability density functions (PDF) of intermittent events often show unimodal structure with ``elevated" tails that deviate from Gaussian predictions. The experimental evidence of the wave-number spectrum characterised by power laws over a wide range of wave-numbers can be directly linked to the PDFs of the underlying turbulent processes described by the values of the L\'{e}vy fractality index $\alpha$. The L\'{e}vy-type turbulent random processes and related anomalous diffusion phenomena have been observed in a wide variety of complex systems such as semiconductors, glassy materials, nano-pores, biological cells, and epidemic spreading. The problem of finding a proper kinetic description for such complex systems is a challenge. The pedagogical applications of simplified models such as the Kuramoto model of random oscillators are particularly helpful in understanding these non-local and non-Gaussian aspects of dynamics in many-body interacting systems. In this work we applied the model to study the impact of both additive and multiplicative non-Gaussian forcing which has not been considered up to now. In particular we are interested to show the dominant impact of singular events with high amplitude on the the long term collective behaviour, and to illustrate the limitations of the Gaussian assumptions in these non-linearly coupled systems. We hope to start a wider discussion on the features that can be expected in the case of strange kinetics and open new fields of application such as that of transport and turbulent dynamics. 

Now we will present the details of the model and the obtained results. At the end of this letter we discuss our findings and draw conclusions.

\section{The numerical set up}
The dynamics of the phases of the oscillators are described by coupled first order differential equations of the form:
\begin{equation}
\dot{\theta}_{j}(t)=\omega_j+(2\pi)^{-1}\sum_{i=1}^{N}J_{ij}sin(\theta_i-\theta_j),\;\;\;\;(j=1,...,N),
\label{theta}
\end{equation}
where $\theta(t)_{j}$ is the phase of the $j$th oscillator with $\dot{\theta}_{j}(t)$ being its time derivative. Here $\omega_j$ is the natural frequency of the oscillator which is assumed to be distributed according to a Gaussian distribution $f(\omega)=exp(-\omega^2/2)/\sqrt 2\pi$. $J_{ij}$ is the strength of the interactions between oscillators $i$th and $j$th and are assumed to be random constants distributed according to an $\alpha$-stable distribution $S(\alpha,\beta,\sigma,\mu)$ with characteristic exponent $0<\alpha\le 2$, skewness $\beta$, scale $\sigma$ and location $\mu$ \cite{Borak,Chambers,Weron,Kirkpatrick,Sakaguchi,Acebron,moradi1,moradi3}. Here we chose $\beta=0$, $\mu=0$, and $\sigma=F/(N\sqrt 2)$ where $F$ is the control parameter as in Ref. \cite{daido2}. Moreover we assume positive and symmetric coupling, i.e. $J_{ij}>0$ and $J_{ij}=J_{ji}$ respectively. The $\alpha$-stable distributions are a general class of distributions which also include Gaussian ($\alpha=2$) and Lorentzian ($\alpha=1$) distributions.

In this work solving Eq. \ref{theta}, the numerical integration is performed using the Runge-Kutta 4th order scheme with time stepping length $\delta t=2\pi\times dt$ with an adaptive time stepping length $dt$ while the sampling time step is $\Delta t=0.01$. The numerical integration is performed for the incoherent initial set with $\theta_j(t=0)$ taken to be positive Gaussian distributed random values for an ensemble of $N$ oscillators. Here we employ an average over a number of different realisations of $J_{ij}$ denoted by $N_{s}=10$, hereafter referred to as ``samples". In the present study, the time span considered is of the order of $2\pi\times 10$. This time span is found to be long enough for the system to reach a steady-state and the numerical noise due to the finite size effects are absent. 

\section{L\'{e}vy coupled oscillators}
\begin{figure}
\includegraphics[width=8.5cm, height=5cm]{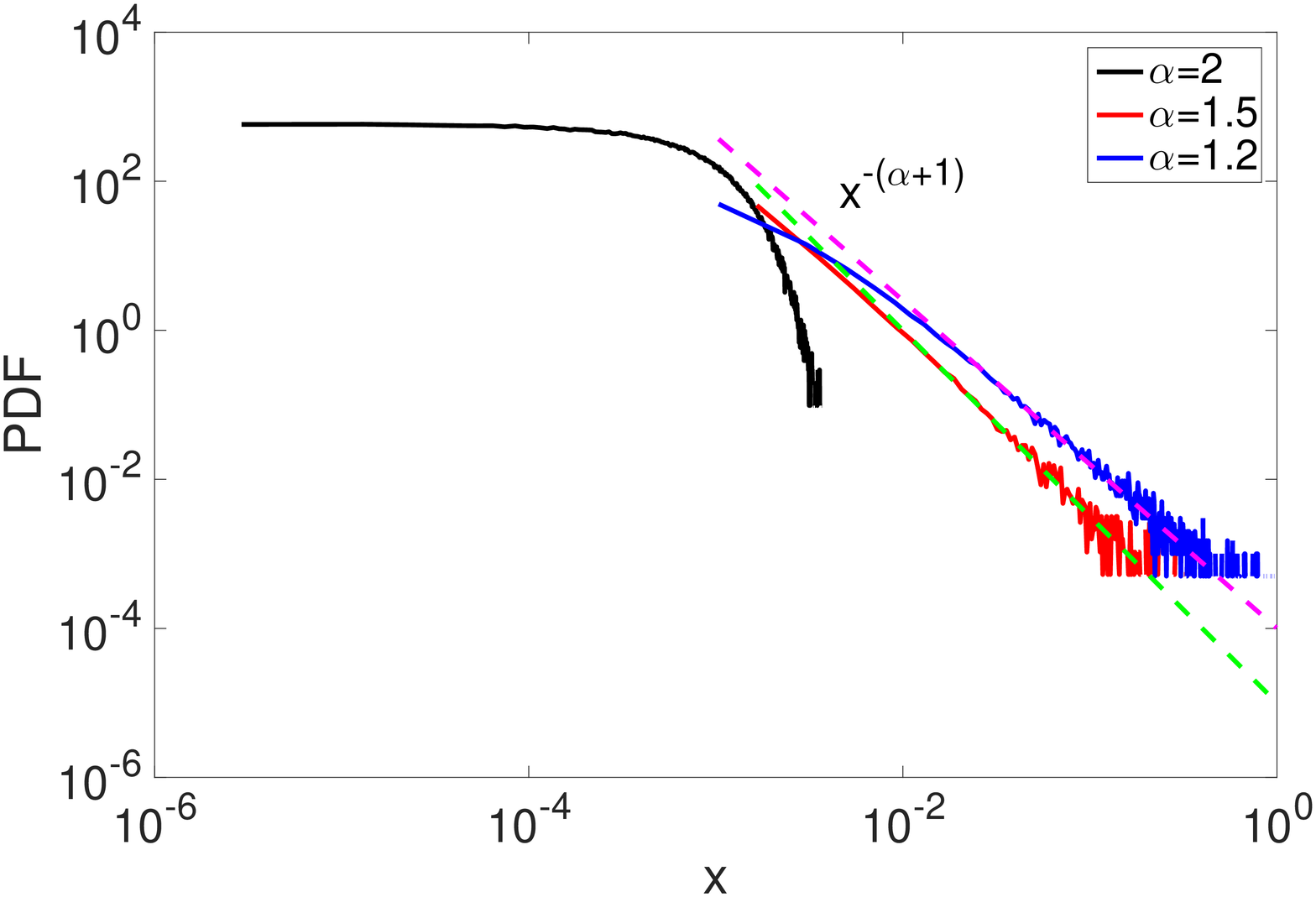}
\caption{\label{fig1} PDFs of the coupling strength parameter $J_{ij}$ according to $\alpha$-stable distribution with $\alpha=2$ (black), $1.5$ (red), $1.2$ (blue). Dotted lines represent the corresponding $x^{-(1+\alpha)}$ fits of the PDFs.}
\end{figure}

We have performed numerical integrations for different values of the fractality index $\alpha$, e.g. $\alpha=2,1.5,1.2$. Figure \ref{fig1} illustrates the PDFs, normalized by their numerical integral, of the coupling strengths $J_{ij}$ for the selected $\alpha$s. Here as we decrease the index $\alpha$ from $2$ to $1.2$ the tails of the distribution become heavier, indicating an increase in the probability of high strength couplings. A $x^{-(1+\alpha)}$ power law decay fit, which is typical of $\alpha$-stable distributions, confirms the proper scaling of the L\'{e}vy stable process generated by the random generator used here, (see Fig. \ref{fig1}).

\begin{figure}
\includegraphics[width=8cm, height=4.5cm]{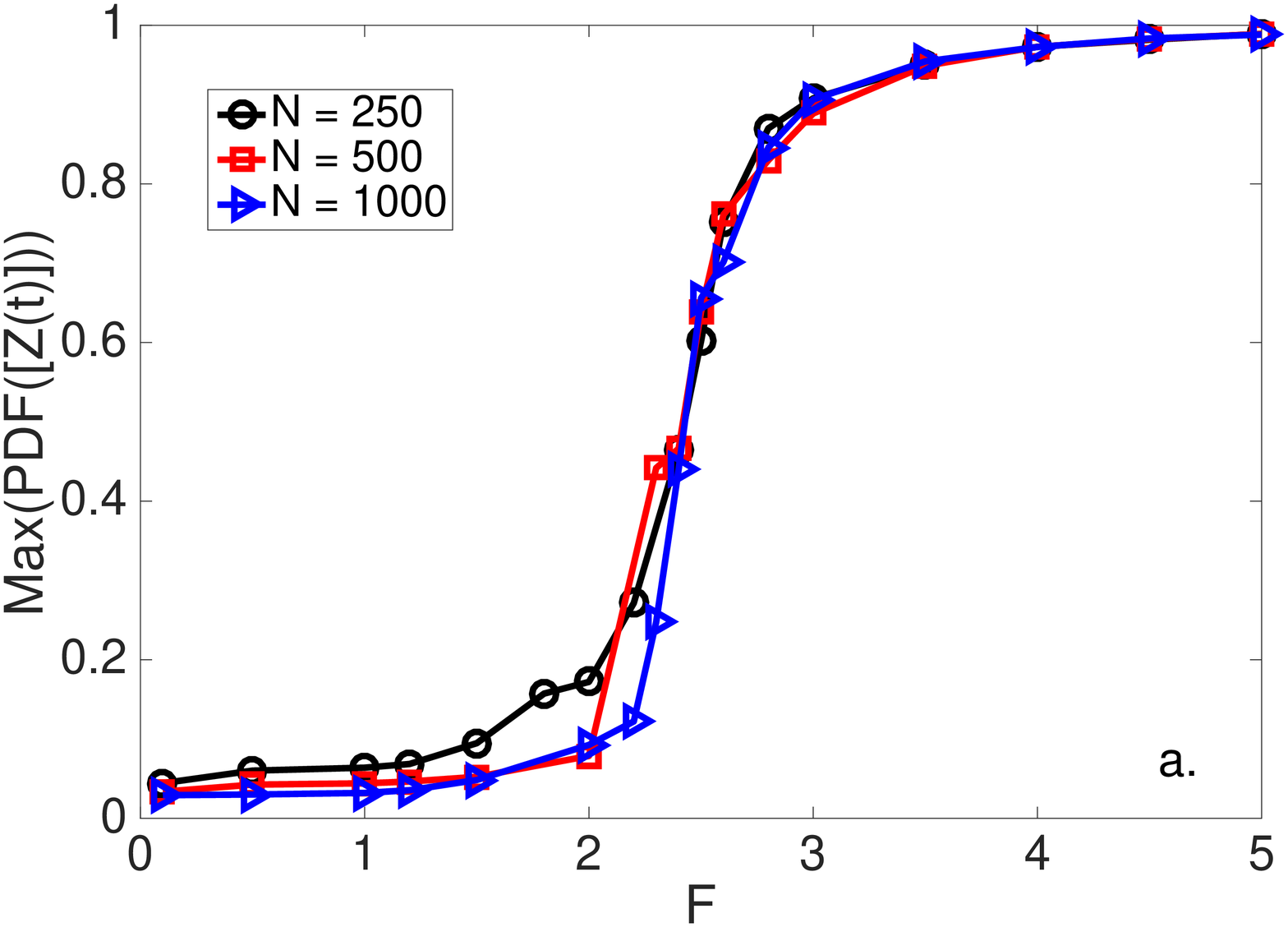}\\
\includegraphics[width=8cm, height=4.5cm]{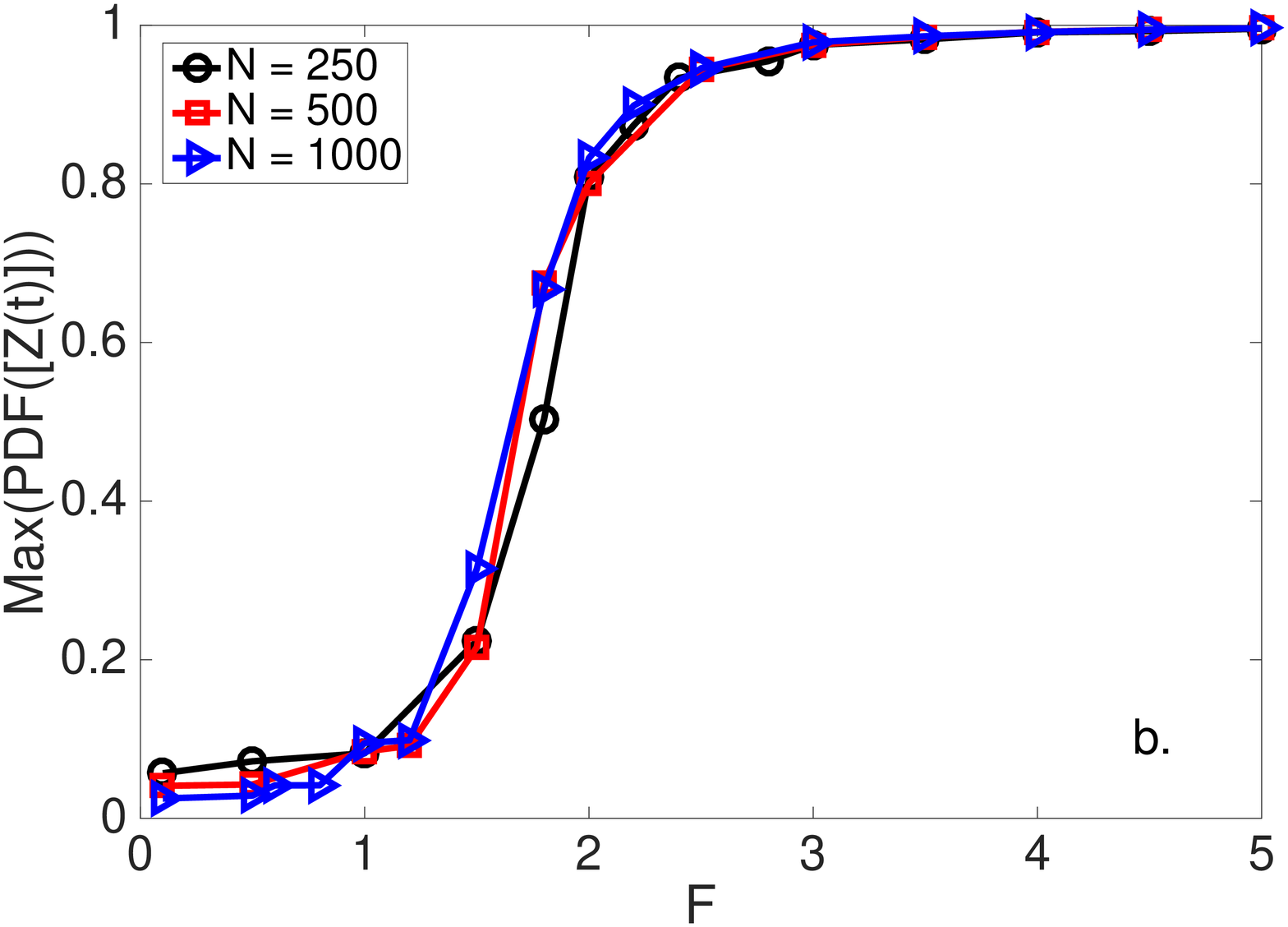}\\
\includegraphics[width=8cm, height=4.5cm]{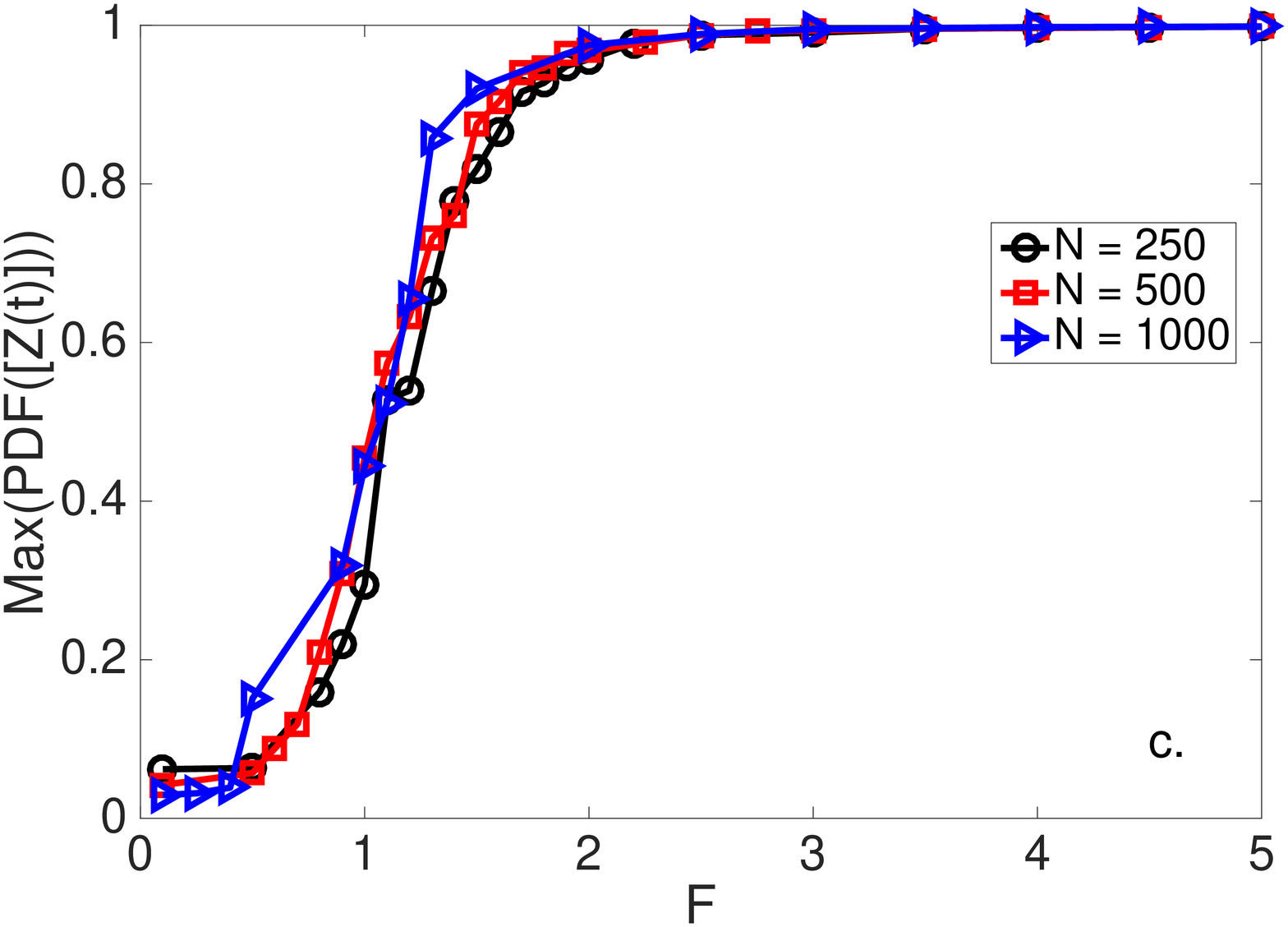}
\caption{\label{fig2} The peaks of the sample averaged $PDF([|Z(t)|])$ as functions of the control parameter $F$ for $\alpha=2$ (a), $\alpha=1.5$ (b) and $\alpha=1.2$ (c). In each figure a scan over different populations are performed with $N=250$ (black line with circles), $N=500$ (red line with squares) and $N=1000$ (blue line with triangles).}
\end{figure}

An analytic expression for the order parameter $Z(t)=\sum_{j=1}^{N} exp(i\theta_{j})/N$ was derived by Kuramoto that describes the synchronisation of the ensemble of oscillators, with $0\le |Z|\le1$. Where $|Z|=0$ corresponds to an asynchronous state while $|Z|=1$ corresponds to a totally synchronous state. We have calculated the values of the order parameter averaged over $N_s$ samples as well as for the various cases considered. Figures \ref{fig2}(a-c) show the order parameter $|Z(t)|$ as a function of $F$ for different numbers of oscillator populations $N = 250,500,1000$, with different values of $\alpha$-stable distribution index $\alpha=2,1.5,1.2$. As can be seen in Fig. \ref{fig2} (a-c) for low values of control parameters i.e. $F \lesssim 2$ the phases are asynchronous with $[|Z(t)|]\approx 0$. As the control parameters increase beyond this threshold the phases bifurcate from an asynchronous to a synchronous state. The threshold where the populations change from a synchronous to asynchronous state, in agreement with the previous reports (see Refs. \cite{daido1, daido2}), is found to be independent of the number of oscillators in the population. In the following we thus, fixed the size of populations to $N=250$.

\begin{figure}
\includegraphics[width=8cm, height=5cm]{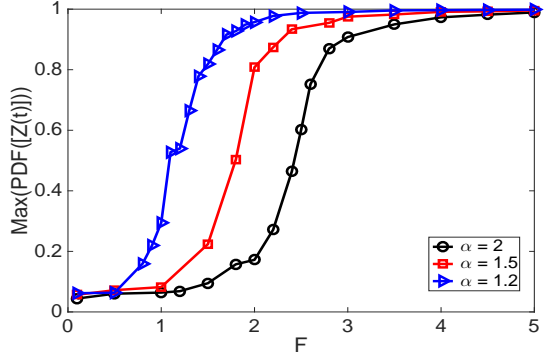}
\caption{\label{fig2.1} The comparison of the maximum of the $PDF([Z(t)])$ as functions of the control parameters $F$ for $\alpha=2$ (black line with circles), $\alpha=1.5$ (red line with squares) and $\alpha=1.2$ (blue line with triangles). Here, N=250.}
\end{figure}

Figure \ref{fig2.1} compares the computed values of the peaks of the PDFs of $[Z(t)]$ vs $F$ for $N = 250$, between the different $\alpha$s. As can be seen in these figures, a bifurcation to a synchronised state occurs as $F$ is increased beyond a critical value $F_c$ which holds for all fractality index values considered in the $\alpha$-stable distribution. However, there is a significant shift to lower values of the criticality parameter $F_c$ as $\alpha$ is decreased from $2$, where $2$ corresponds to the Gaussian distribution. This indicates that the extreme events from the tails of the distribution can provide a faster synchronisation of the coupled oscillators, and as the tail gets heavier by moving from $\alpha=1.5$ to $1.2$, $F_{dc}$ is shifted to even lower values. 

\begin{figure}
\includegraphics[width=8cm, height=5cm]{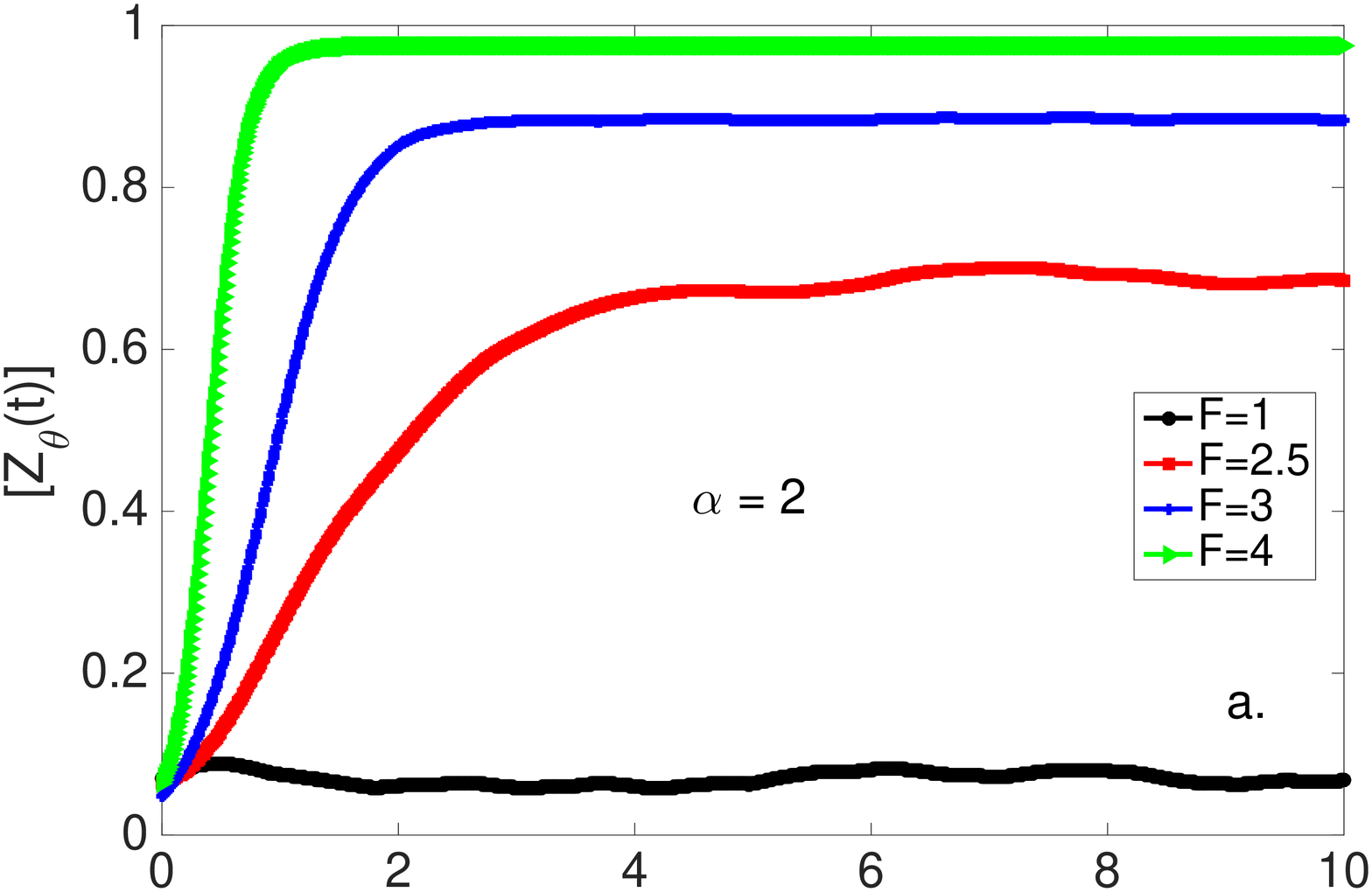}\\
\includegraphics[width=8cm, height=5cm]{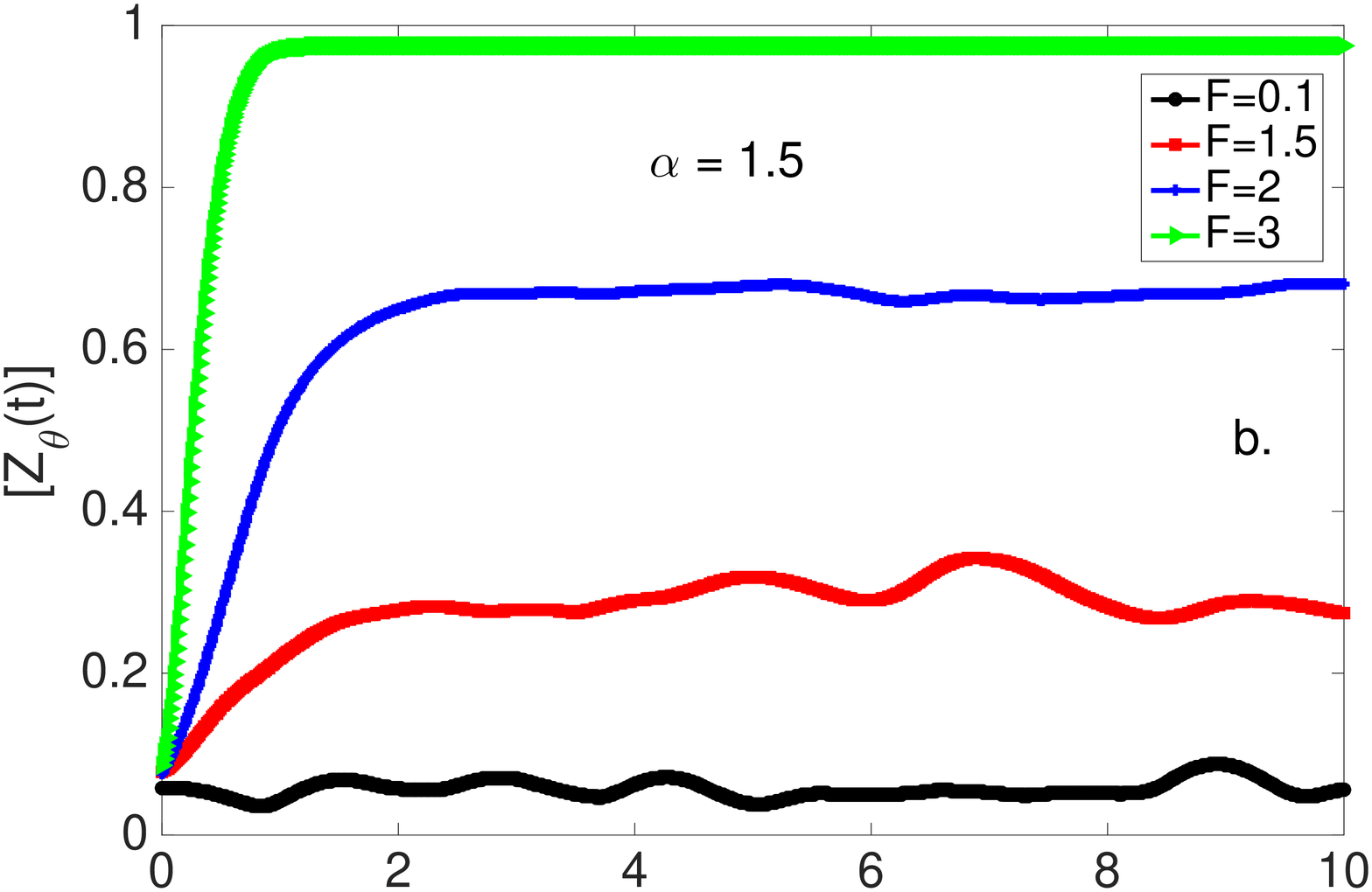}\\
\includegraphics[width=8cm, height=5cm]{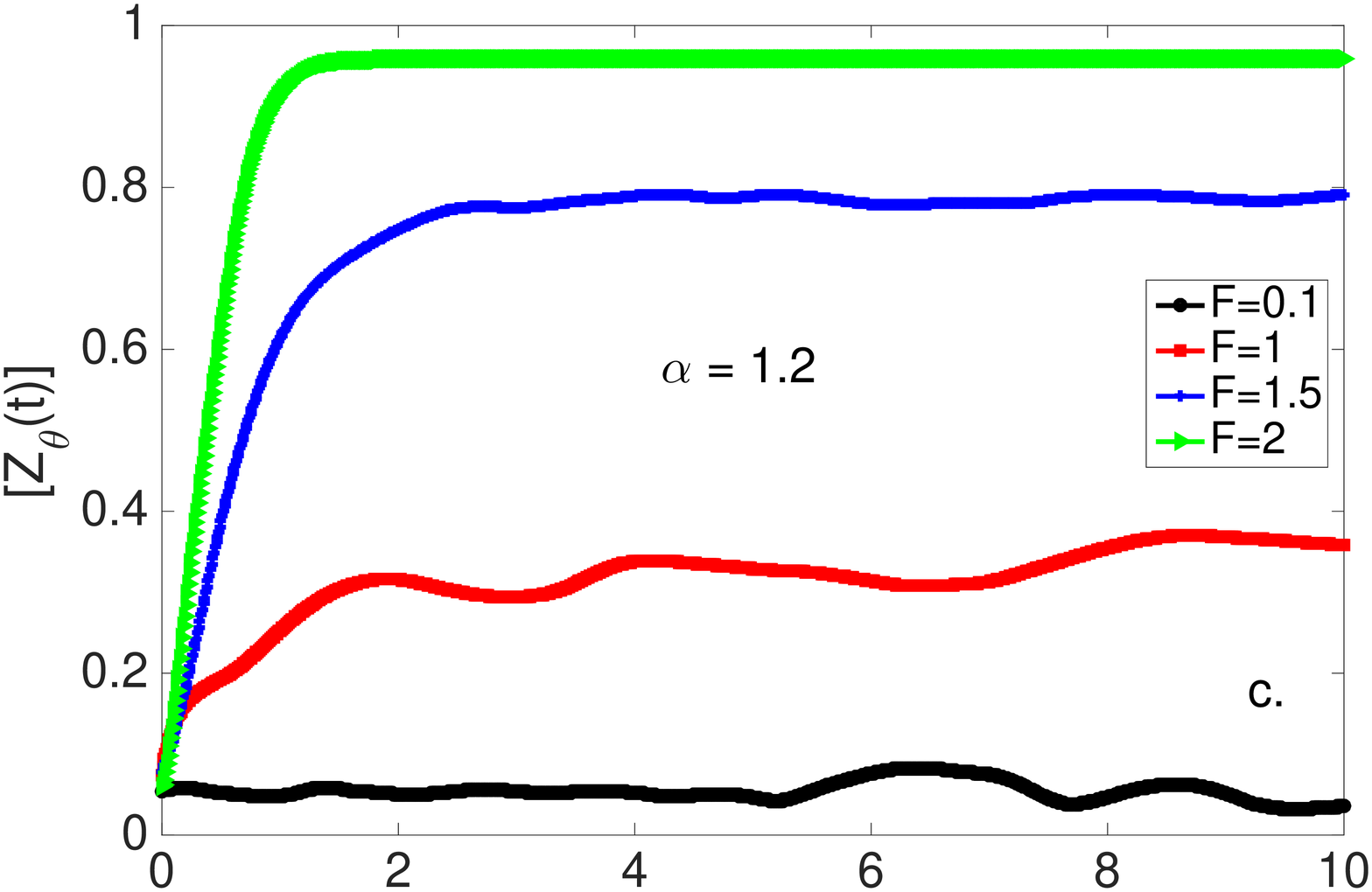}
\caption{\label{fig3} Sample averaged $[|Z(t)|]$ for $\alpha=2$ (a), $\alpha=1.5$ (b) and $\alpha=1.2$ (c) as functions of simulation time $t$, and for different values of control parameter $F$.}
\end{figure}

\begin{figure}
\includegraphics[width=8cm, height=5cm]{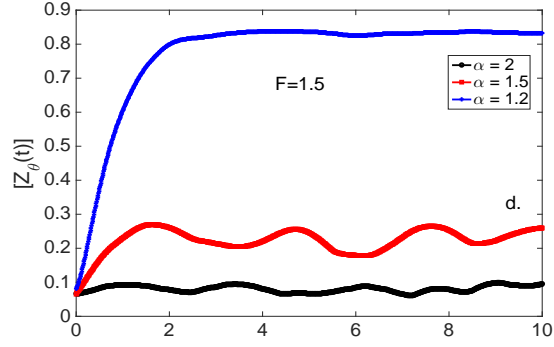}
\caption{\label{fig3.1} The comparison of the sample averaged $[|Z(t)|]$ as functions of simulation time $t$, for $\alpha=2$ (black line ), $\alpha=1.5$ (red line) and $\alpha=1.2$ (blue line). Here we have fixed the order parameter $F=1.5$ in all cases.}
\end{figure}

Figures \ref{fig3}(a-c) sample averaged $[|Z(t)|]$ as functions of time for different control parameters and $\alpha$ index. Also, here we observe evolution of the system of oscillators from asynchronous initial state towards fully synchronised state as the control parameter is increased. A comparison between the different $\alpha$ cases and at the same level of $F= 1.5$ is shown in Fig. \ref{fig3.1}. Here a remarkably interesting result is that a coupled system of oscillators following a non-Gaussian random coupling can reach a synchronised state in considerably shorter time period than for case of a Gaussian random coupling. 

\begin{figure}
\includegraphics[width=8cm, height=5cm]{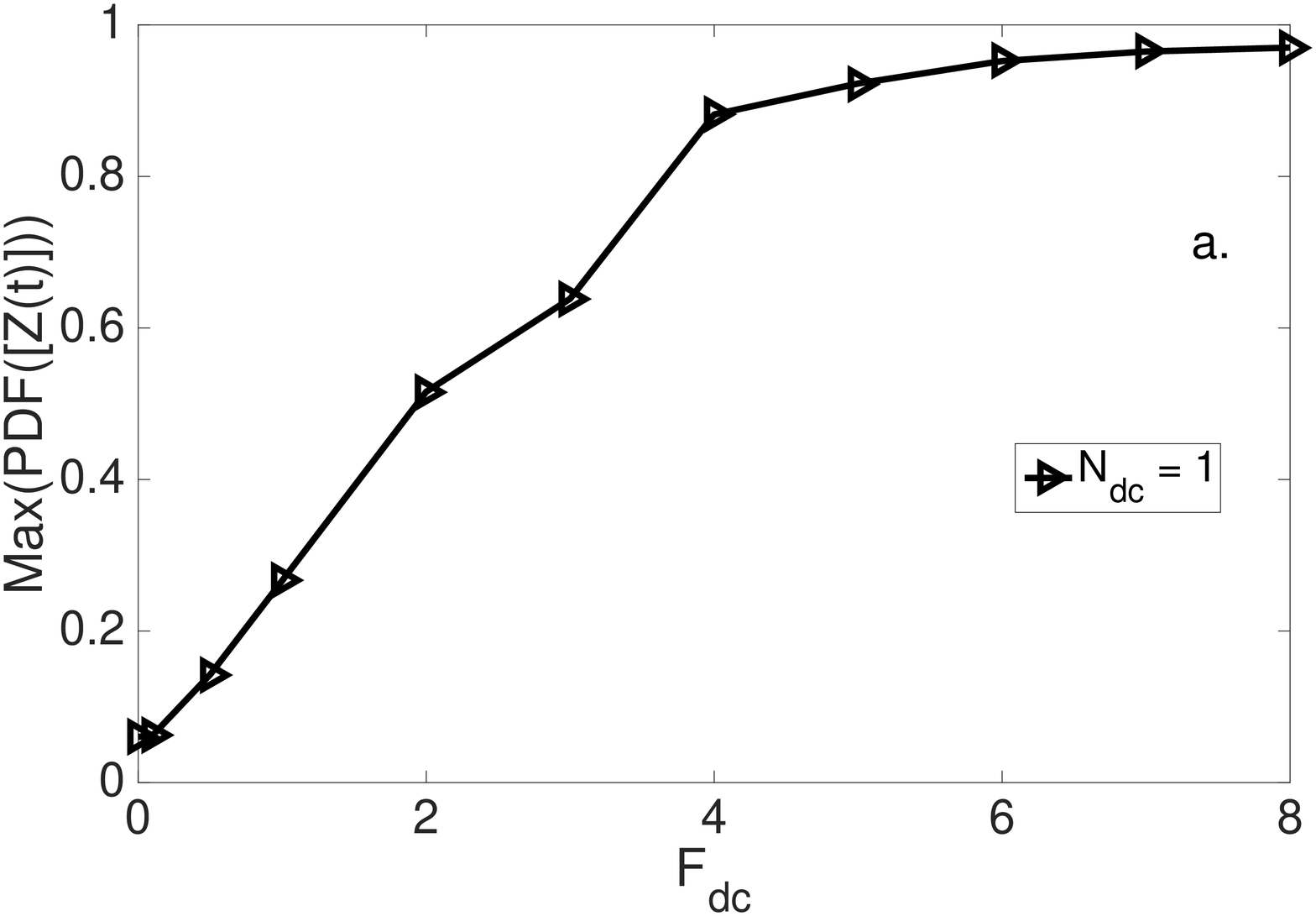}\\
\includegraphics[width=8cm, height=5cm]{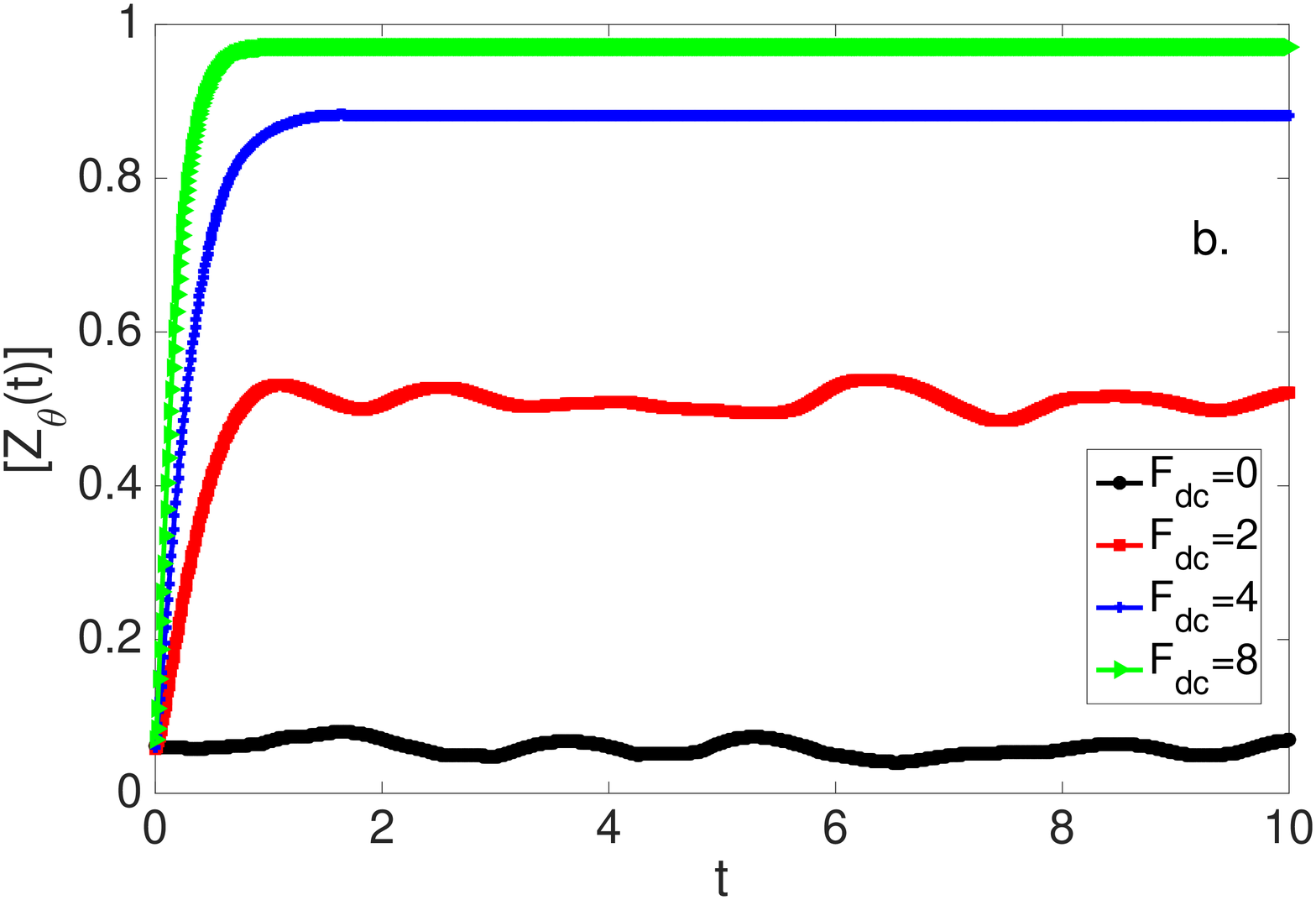}
\caption{\label{fig4} (a) The maximum of the $PDF([Z(t)])$ as a function of the control parameter $F_{dc}$ for $N_{dc=1}$. Here the rest of the oscillators are coupled through Gaussian distributed $J_{ij}$ with $F=0.1$. (b) The comparison of the sample averaged $[|Z(t)|]$ as functions of simulation time $t$, for $F_{dc}=0$ (black line ), $F_{dc}=2$ (red line), $F_{dc}=4$ (blue line) and $F_{dc}=6$ (green line). Here $N_{dc}=1$.}
\end{figure}

\begin{figure}
\includegraphics[width=5cm, height=3cm]{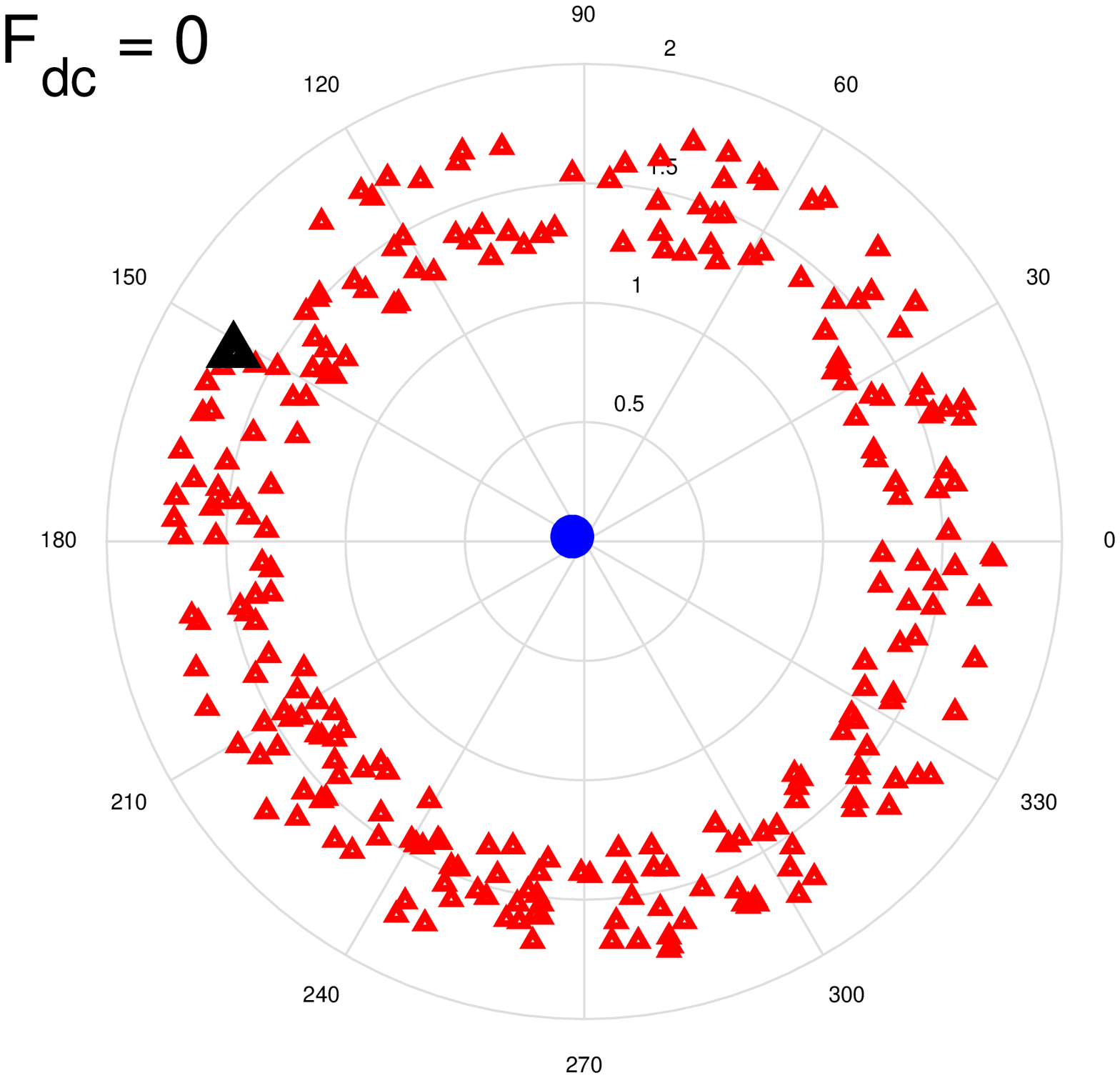}\\
\includegraphics[width=5cm, height=3cm]{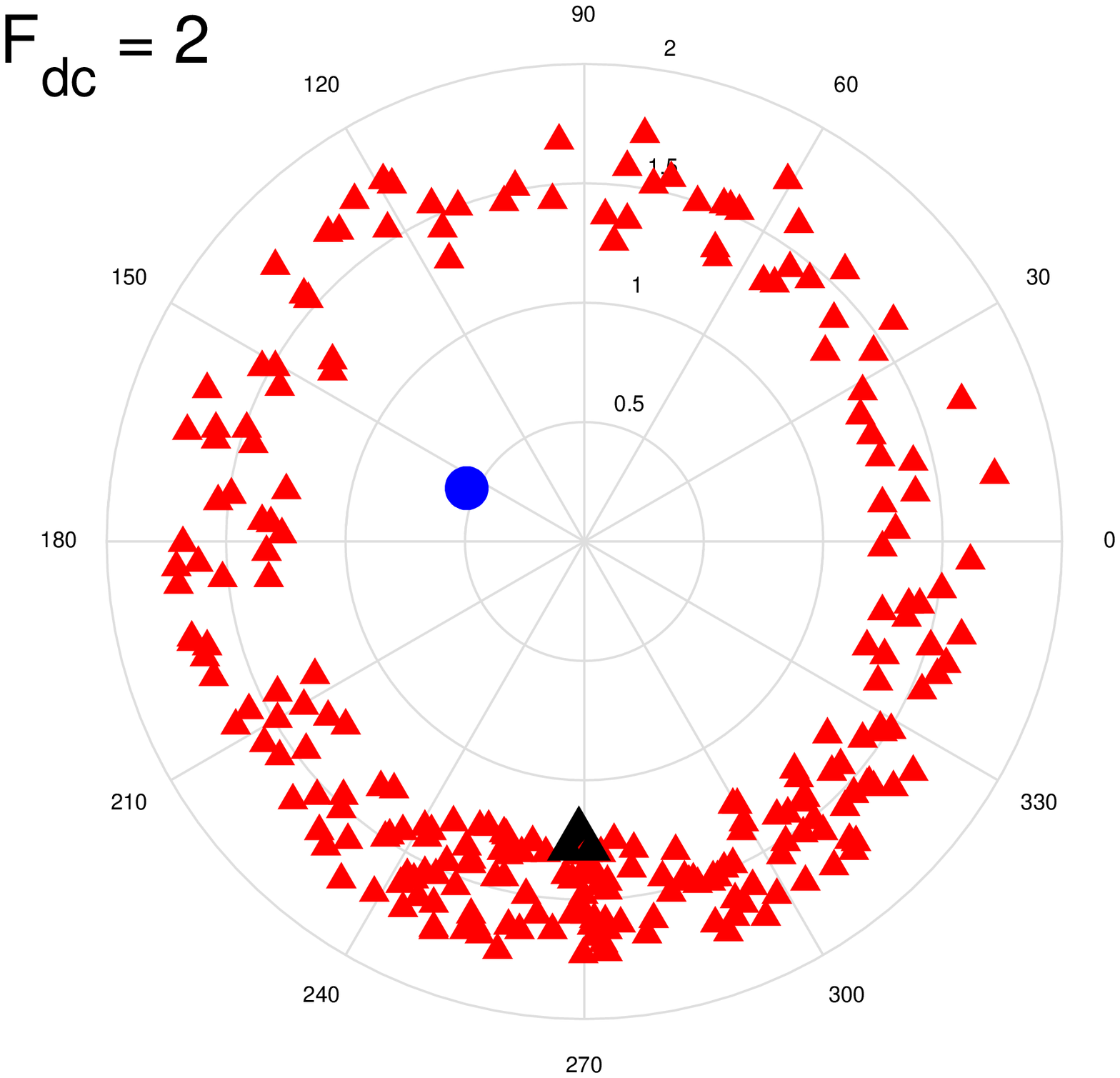}\\
\includegraphics[width=5cm, height=3cm]{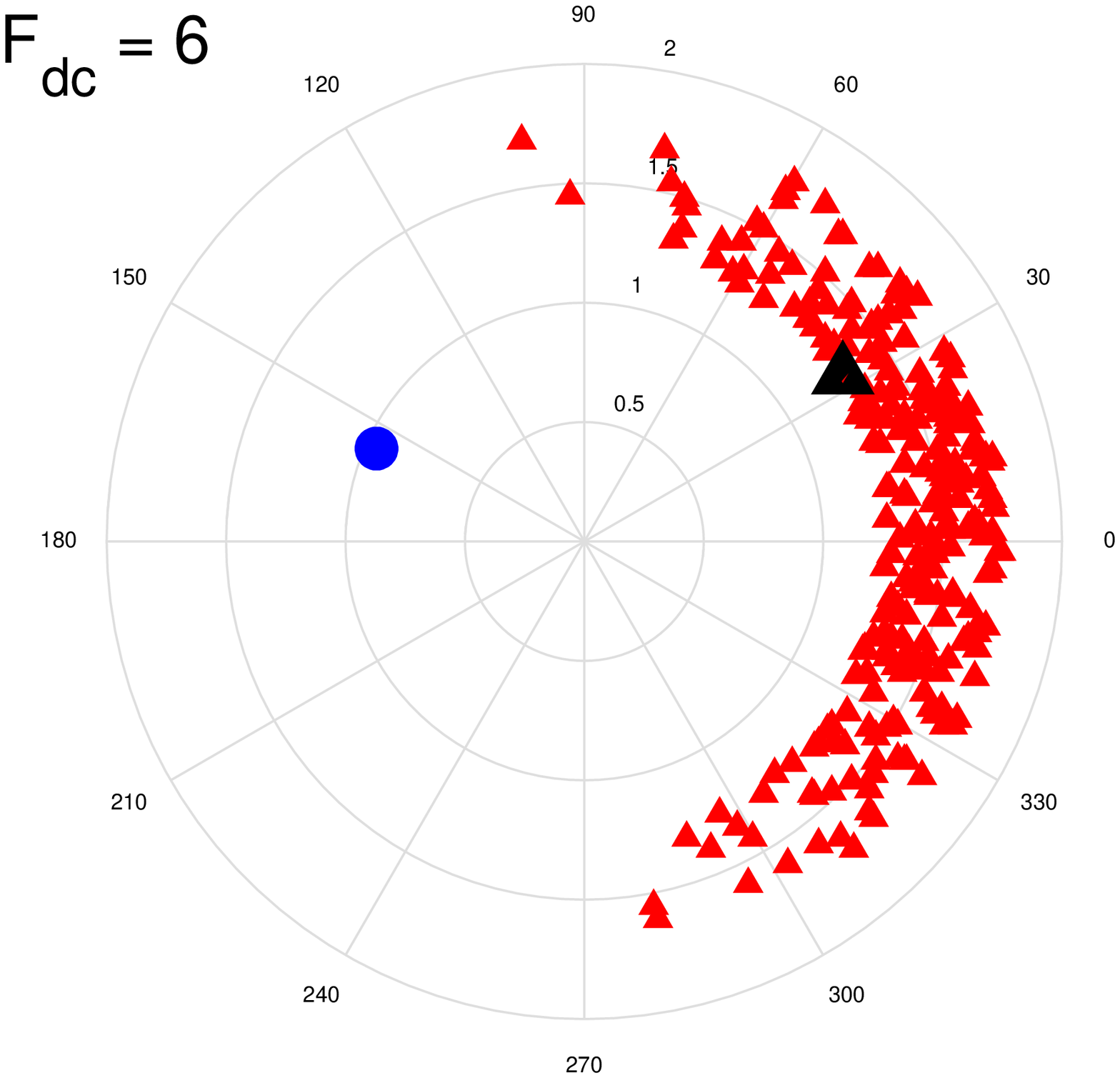}
\caption{\label{fig5} Polar representation of the phases at the end of the simulation time for various levels of $F_{dc}=0,2, 6$ from top to bottom respectively. The black triangle represent the outlier oscillator and the red circle illustrates the value of the sample averaged order parameter $[|Z(t)|]$.}
\end{figure}

\begin{figure}
\includegraphics[width=8cm, height=5cm]{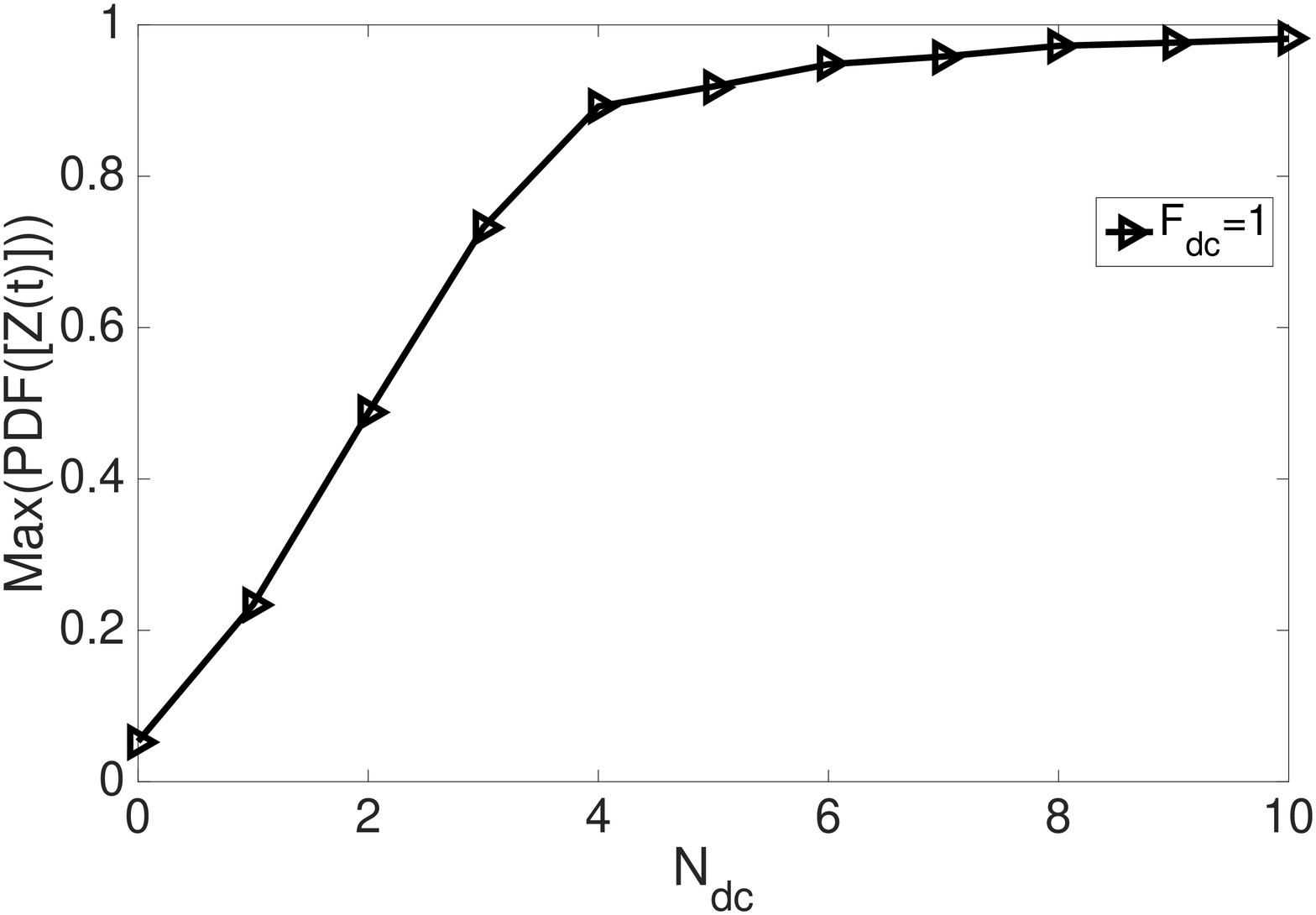}
\caption{\label{fig6} The maximum of the $PDF([Z(t)])$ as a function of the $N_{dc}$ and with fixed value of $F_{dc}=1$.}
\end{figure}

Following our findings discussed above we examined the role of extreme events in the tails of L\'{e}vy distributions by artificially introducing such events into the coupled system. Starting with the set of parameters needed to create an a-synchronized state e.g. control parameter of $J=0.1$ with $J_{ij}$s following Gaussian distribution (see Fig. \ref{fig4}); we introduced $N_{dc}$ (different coupling) additional oscillators with a given constant coupling strength of $F_{dc}$ and investigated the minimum number of oscillators needed to generate a self-synchronised coupled state. As shown in Figs. \ref{fig4}(a and b) we find that for these values, even one strongly coupled oscillator i.e. extreme event, is enough to create synchronisation in the system. The respective polar view representation of the phases of the oscillators are shown in Fig. \ref{fig5} for three levels of $F_{dc} =0,2, 6$. As shown in this figure, the synchronisation here, is due to the strong impact of the outlier event on the mean field which results in the synchronisation of even weakly coupled oscillators. Also, for a given value of $F_{dc}=1$ corresponding to moderate coupling strength between oscillators, as we increase the number of such oscillators, i.e. $N_{dc}$, we can reach a synchronised state for as low values of $N_{dc}$ as $N_{dc} =10$, see Fig. \ref{fig6}. In the next section, we examine the impact of an external L\'{e}vy distributed noise on the synchronisation properties of the oscillators. The stochastic differential equations of \ref{theta2} and \ref{theta3} are solved using a Euler-Maruyama method \cite{Higham}. Stochastic Kuramoto models with disorder either in coupling strength or the extra noise have been studied previously \cite{daido2,Kirkpatrick,sonnenschein,Sakaguchi,Acebron}, in particularly in relation to XY spin-glasses where $\omega_i=0$ is introduced to examine the behaviour of frustrated magnets \cite{Kirkpatrick}.

\paragraph*{L\'{e}vy coupled oscillators with external noise}
We introduced an additional source of disorder into the system of coupled oscillators following:
\begin{equation}
\dot{\theta}_{j}(t)=\omega_j+(2\pi)^{-1}\sum_{i=1}^{N}J_{ij}sin(\theta_i-\theta_j) + \eta_{j},\;\;\;\;(j=1,...,N),
\label{theta2}
\end{equation}
where
\begin{equation}
 \eta_{j} = \sum_{dt}\delta t^{(1/\alpha)} W(t),\;\;\;\;(j=1,...,N),
\label{theta3}
\end{equation}

Here $W(t)$ are random values distributed according to an $\alpha$-stable distribution with $0<\alpha\le 2$, $\beta=0$, $\mu=0$, and $\sigma=\gamma/\sqrt 2$. $\gamma$ represents the control parameter of this external noise. The summation over $\delta t $ where $dt =n \delta t$ represents the accumulation of various non-linear processes over $n$ time steps $\delta t$, and can take both positive and negative values. We have included both the possible sources of disorder as well as considered the effect of L\'{e}vy noise on the synchronisation phenomena.

\begin{figure}
\includegraphics[width=8cm, height=5cm]{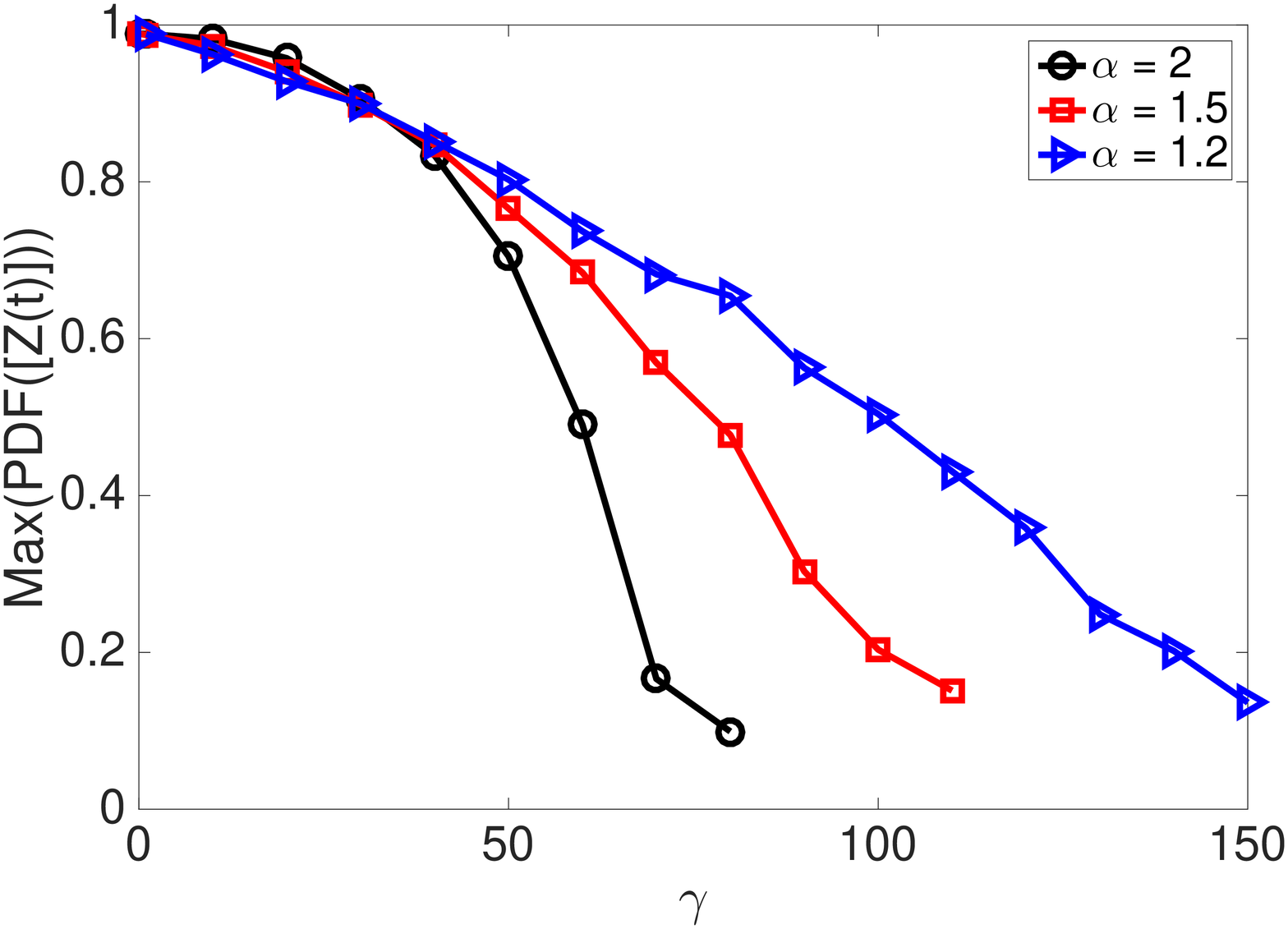}
\caption{\label{fig7} The maximum of the $PDF([Z(t)])$ as a function of the $\gamma$ for $\alpha=2$ (black line with circles), $\alpha=1.5$ (red line with squares) and $\alpha=1.2$ (blue line with triangles). Here $N=250$, $F=3$, $dt=0.01$, and $n=1$.}
\end{figure}

Following the results of the previous section, we start the system by setting the control parameter $F=3$ for which the system, without the external force, is synchronised. Here, we assume $J_{ij}$ are Gaussian random values. The dependence of the maximum of the $PDF([Z(t)])$ on the control parameter $\gamma$ for the cases with $\alpha =2$, $1.5$ and $1.2$ are shown in Fig. \ref{fig7}, where in all cases, we find that as $\gamma$ is increased $PDF([Z(t)])$ decays to $0$ indicating that the synchronised system of coupled oscillators moves towards an a-synchronised state. As the two last terms of the right hand side of the eq. \ref{theta2} compete to bring the system to synchrony or asynchrony, correspondingly, it is not surprising that as the strength of the force is increased the competition is considerably more pronounced by the a-synchronising effect of the external force. We observe that the $\gamma$ threshold for the a-synchronization of the coupled state is significantly increased as we change from a Gaussian type external force with index $\alpha=2$ to L\'{e}vy type noise with index $\alpha=1.5$ and $1.2$. The reason for this up-shift lies in the pre-factor $\delta t^{(1/\alpha)}$ which reduces as $\alpha$ is decreased. Thus, on the one hand, as the system has a strong non-linear coupling, the rare events in the stochastic force do not influence the collective system strongly, and on the other hand, the impact of the external L\'{e}vy distribution force is reduced as compare to Gaussian distributed forces due to pre-factor $\delta t^{(1/\alpha)}$. Similar results were obtained for cases with $n=1, 4, 8,$ and $16$.

\section{Summary} We have studied self-organization properties within the paradigm of Kuramoto oscillators by extending the stochastic background to accommodate L\'{e}vy stable processes. We present results for two different and distinct versions based on the first order system of differential equations. We have found that synchronization is severely dependent on the fractality index ($1 < \alpha \le 2.0$) of the L\'{e}vy stable process. By decreasing $\alpha$, synchronization is found, in much the same way as it occurs for a Gaussian process, however at significantly lower levels of the control parameter. By including an external stochastic force term, representing additional unknown properties of a non-linear dynamical system, we find that a synchronised state is desynchronised as the strength of the force is increased. However, for the forces following L\'{e}vy distributions with index $\alpha =1.5$ and $1.2$, there is a significant up-shift in the strength of the external force for which the oscillators are desynchronised. Note that in this case the strength of the forcing can drive the system out of synchrony. In conclusion, we find that extreme events can govern the long term behaviour of non-linear systems, such as the Kuramoto model. In order to build a realistic model of such systems for the purpose of making long term predictions the impact of extreme events have to be analysed and carefully characterized. Overly simplistic descriptions of the statistics of the underlying fluctuations can lead to misrepresentations of the long term behaviour of the dynamical system.

\end{document}